\documentstyle[12pt]{article}

\def\PsfigVersion{1.10}
\def\setDriver{\DvipsDriver} 
\ifx\undefined\psfig\else\endinput\fi
\let\LaTeXAtSign=\@
\let\@=\relax
\edef\psfigRestoreAt{\catcode`\@=\number\catcode`@\relax}
\catcode`\@=11\relax
\newwrite\@unused
\def\ps@typeout#1{{\let\protect\string\immediate\write\@unused{#1}}}

\def\DvipsDriver{
	\ps@typeout{psfig/tex \PsfigVersion -dvips}
\def\PsfigSpecials{\DvipsSpecials} 	\def\ps@dir{/}
\def\ps@predir{} }
\def\OzTeXDriver{
	\ps@typeout{psfig/tex \PsfigVersion -oztex}
	\def\PsfigSpecials{\OzTeXSpecials}
	\def\ps@dir{:}
	\def\ps@predir{:}
	\catcode`\^^J=5
}

\def\figurepath{./:}

\def\DoPaths#1{\expandafter\EachPath#1\stoplist}
\def\leer{}
\def\EachPath#1:#2\stoplist{
  \ExistsFile{#1}{\SearchedFile}
  \ifx#2\leer
  \else
    \expandafter\EachPath#2\stoplist
  \fi}
\def\ps@dir{/}
\def\ExistsFile#1#2{%
   \openin1=\ps@predir#1\ps@dir#2
   \ifeof1
       \closein1
   \else
       \closein1
        \ifx\ps@founddir\leer
           \edef\ps@founddir{#1}
        \fi
   \fi}
\def\get@dir#1{%
  \def\ps@founddir{}
  \def\SearchedFile{#1}
  \DoPaths\figurepath
}

\def\@nnil{\@nil}
\def\@empty{}
\def\@psdonoop#1\@@#2#3{}
\def\@psdo#1:=#2\do#3{\edef\@psdotmp{#2}\ifx\@psdotmp\@empty \else
    \expandafter\@psdoloop#2,\@nil,\@nil\@@#1{#3}\fi}
\def\@psdoloop#1,#2,#3\@@#4#5{\def#4{#1}\ifx #4\@nnil \else
       #5\def#4{#2}\ifx #4\@nnil \else#5\@ipsdoloop #3\@@#4{#5}\fi\fi}
\def\@ipsdoloop#1,#2\@@#3#4{\def#3{#1}\ifx #3\@nnil 
       \let\@nextwhile=\@psdonoop \else
      #4\relax\let\@nextwhile=\@ipsdoloop\fi\@nextwhile#2\@@#3{#4}}
\def\@tpsdo#1:=#2\do#3{\xdef\@psdotmp{#2}\ifx\@psdotmp\@empty \else
    \@tpsdoloop#2\@nil\@nil\@@#1{#3}\fi}
\def\@tpsdoloop#1#2\@@#3#4{\def#3{#1}\ifx #3\@nnil 
       \let\@nextwhile=\@psdonoop \else
      #4\relax\let\@nextwhile=\@tpsdoloop\fi\@nextwhile#2\@@#3{#4}}
\ifx\undefined\fbox
\newdimen\fboxrule
\newdimen\fboxsep
\newdimen\ps@tempdima
\newbox\ps@tempboxa
\long\def\fbox#1{\leavevmode\setbox\ps@tempboxa\hbox{#1}\ps@tempdima\fboxrule
    \advance\ps@tempdima \fboxsep \advance\ps@tempdima \dp\ps@tempboxa
   \hbox{\lower \ps@tempdima\hbox
  {\vbox{\hrule height \fboxrule
          \hbox{\vrule width \fboxrule \hskip\fboxsep
          \vbox{\vskip\fboxsep \box\ps@tempboxa\vskip\fboxsep}\hskip 
                 \fboxsep\vrule width \fboxrule}
                 \hrule height \fboxrule}}}}
\fi
\newread\ps@stream
\newif\ifnot@eof       
\newif\if@noisy        
\newif\if@atend        
\newif\if@psfile       
{\catcode`\%=12\global\gdef\epsf@start{
\def\epsf@PS{PS}
\def\epsf@getbb#1{%
\openin\ps@stream=\ps@predir#1
\ifeof\ps@stream\ps@typeout{Error, File #1 not found}\else
   {\not@eoftrue \chardef\other=12
    \def\do##1{\catcode`##1=\other}\dospecials \catcode`\ =10
    \loop
       \if@psfile
	  \read\ps@stream to \epsf@fileline
       \else{
	  \obeyspaces
          \read\ps@stream to \epsf@tmp\global\let\epsf@fileline\epsf@tmp}
       \fi
       \ifeof\ps@stream\not@eoffalse\else
       \if@psfile\else
       \expandafter\epsf@test\epsf@fileline:. \\%
       \fi
          \expandafter\epsf@aux\epsf@fileline:. \\%
       \fi
   \ifnot@eof\repeat
   }\closein\ps@stream\fi}%
\long\def\epsf@test#1#2#3:#4\\{\def\epsf@testit{#1#2}
			\ifx\epsf@testit\epsf@start\else
\ps@typeout{Warning! File does not start with `\epsf@start'.  It may not be a PostScript file.}
			\fi
			\@psfiletrue} 
{\catcode`\%=12\global\let\epsf@percent=
\long\def\epsf@aux#1#2:#3\\{\ifx#1\epsf@percent
   \def\epsf@testit{#2}\ifx\epsf@testit\epsf@bblit
	\@atendfalse
        \epsf@atend #3 . \\%
	\if@atend	
	   \if@verbose{
		\ps@typeout{psfig: found `(atend)'; continuing search}
	   }\fi
        \else
        \epsf@grab #3 . . . \\%
        \not@eoffalse
        \global\no@bbfalse
        \fi
   \fi\fi}%
\def\epsf@grab #1 #2 #3 #4 #5\\{%
   \global\def\epsf@llx{#1}\ifx\epsf@llx\empty
      \epsf@grab #2 #3 #4 #5 .\\\else
   \global\def\epsf@lly{#2}%
   \global\def\epsf@urx{#3}\global\def\epsf@ury{#4}\fi}%
\def\epsf@atendlit{(atend)} 
\def\epsf@atend #1 #2 #3\\{%
   \def\epsf@tmp{#1}\ifx\epsf@tmp\empty
      \epsf@atend #2 #3 .\\\else
   \ifx\epsf@tmp\epsf@atendlit\@atendtrue\fi\fi}

\chardef\psletter = 11 
\chardef\other = 12

\newif \ifdebug 
\newif\ifc@mpute 
\c@mputetrue 

\let\then = \relax
\def\r@dian{pt }
\let\r@dians = \r@dian
\let\dimensionless@nit = \r@dian
\let\dimensionless@nits = \dimensionless@nit
\def\internal@nit{sp }
\let\internal@nits = \internal@nit
\newif\ifstillc@nverging
\def \Mess@ge #1{\ifdebug \then \message {#1} \fi}

{ 
	\catcode `\@ = \psletter
	\gdef \nodimen {\expandafter \n@dimen \the \dimen}
	\gdef \term #1 #2 #3%
	       {\edef \t@ {\the #1}
		\edef \t@@ {\expandafter \n@dimen \the #2\r@dian}%
		\t@rm {\t@} {\t@@} {#3}%
	       }
	\gdef \t@rm #1 #2 #3%
	       {{%
		\count 0 = 0
		\dimen 0 = 1 \dimensionless@nit
		\dimen 2 = #2\relax
		\Mess@ge {Calculating term #1 of \nodimen 2}%
		\loop
		\ifnum	\count 0 < #1
		\then	\advance \count 0 by 1
			\Mess@ge {Iteration \the \count 0 \space}%
			\Multiply \dimen 0 by {\dimen 2}%
			\Mess@ge {After multiplication, term = \nodimen 0}%
			\Divide \dimen 0 by {\count 0}%
			\Mess@ge {After division, term = \nodimen 0}%
		\repeat
		\Mess@ge {Final value for term #1 of 
				\nodimen 2 \space is \nodimen 0}%
		\xdef \Term {#3 = \nodimen 0 \r@dians}%
		\aftergroup \Term
	       }}
	\catcode `\p = \other
	\catcode `\t = \other
	\gdef \n@dimen #1pt{#1} 
}

\def \Divide #1by #2{\divide #1 by #2} 

\def \Multiply #1by #2
       {{
	\count 0 = #1\relax
	\count 2 = #2\relax
	\count 4 = 65536
	\Mess@ge {Before scaling, count 0 = \the \count 0 \space and
			count 2 = \the \count 2}%
	\ifnum	\count 0 > 32767 
	\then	\divide \count 0 by 4
		\divide \count 4 by 4
	\else	\ifnum	\count 0 < -32767
		\then	\divide \count 0 by 4
			\divide \count 4 by 4
		\else
		\fi
	\fi
	\ifnum	\count 2 > 32767 
	\then	\divide \count 2 by 4
		\divide \count 4 by 4
	\else	\ifnum	\count 2 < -32767
		\then	\divide \count 2 by 4
			\divide \count 4 by 4
		\else
		\fi
	\fi
	\multiply \count 0 by \count 2
	\divide \count 0 by \count 4
	\xdef \product {#1 = \the \count 0 \internal@nits}%
	\aftergroup \product
       }}

\def\r@duce{\ifdim\dimen0 > 90\r@dian \then   
		\multiply\dimen0 by -1
		\advance\dimen0 by 180\r@dian
		\r@duce
	    \else \ifdim\dimen0 < -90\r@dian \then  
		\advance\dimen0 by 360\r@dian
		\r@duce
		\fi
	    \fi}

\def\Sine#1%
       {{%
	\dimen 0 = #1 \r@dian
	\r@duce
	\ifdim\dimen0 = -90\r@dian \then
	   \dimen4 = -1\r@dian
	   \c@mputefalse
	\fi
	\ifdim\dimen0 = 90\r@dian \then
	   \dimen4 = 1\r@dian
	   \c@mputefalse
	\fi
	\ifdim\dimen0 = 0\r@dian \then
	   \dimen4 = 0\r@dian
	   \c@mputefalse
	\fi
	\ifc@mpute \then
		\divide\dimen0 by 180
		\dimen0=3.141592654\dimen0
		\dimen 2 = 3.1415926535897963\r@dian 
		\divide\dimen 2 by 2 
		\Mess@ge {Sin: calculating Sin of \nodimen 0}%
		\count 0 = 1 
		\dimen 2 = 1 \r@dian 
		\dimen 4 = 0 \r@dian 
		\loop
			\ifnum	\dimen 2 = 0 
			\then	\stillc@nvergingfalse 
			\else	\stillc@nvergingtrue
			\fi
			\ifstillc@nverging 
			\then	\term {\count 0} {\dimen 0} {\dimen 2}%
				\advance \count 0 by 2
				\count 2 = \count 0
				\divide \count 2 by 2
				\ifodd	\count 2 
				\then	\advance \dimen 4 by \dimen 2
				\else	\advance \dimen 4 by -\dimen 2
				\fi
		\repeat
	\fi		
			\xdef \sine {\nodimen 4}%
       }}

\def\Cosine#1{\ifx\sine\UnDefined\edef\Savesine{\relax}\else
		             \edef\Savesine{\sine}\fi
	{\dimen0=#1\r@dian\advance\dimen0 by 90\r@dian
	 \Sine{\nodimen 0}
	 \xdef\cosine{\sine}
	 \xdef\sine{\Savesine}}}	      

\def\psdraft{
	\def\@psdraft{0}
}
\def\psfull{
	\def\@psdraft{100}
}

\psfull

\newif\if@scalefirst
\def\psscalefirst{\@scalefirsttrue}
\def\psrotatefirst{\@scalefirstfalse}
\psrotatefirst

\newif\if@draftbox
\def\psnodraftbox{
	\@draftboxfalse
}
\def\psdraftbox{
	\@draftboxtrue
}
\@draftboxtrue

\newif\if@prologfile
\newif\if@postlogfile
\def\pssilent{
	\@noisyfalse
}
\def\psnoisy{
	\@noisytrue
}
\psnoisy
\newif\if@bbllx
\newif\if@bblly
\newif\if@bburx
\newif\if@bbury
\newif\if@height
\newif\if@width
\newif\if@rheight
\newif\if@rwidth
\newif\if@angle
\newif\if@clip
\newif\if@verbose
\def\@p@@sclip#1{\@cliptrue}
\newif\if@decmpr
\def\@p@@sfigure#1{\def\@p@sfile{null}\def\@p@sbbfile{null}\@decmprfalse
   \openin1=\ps@predir#1
   \ifeof1
	\closein1
	\get@dir{#1}
	\ifx\ps@founddir\leer
		\openin1=\ps@predir#1.bb
		\ifeof1
			\closein1
			\get@dir{#1.bb}
			\ifx\ps@founddir\leer
				\ps@typeout{Can't find #1 in \figurepath}
			\else
				\@decmprtrue
				\def\@p@sfile{\ps@founddir\ps@dir#1}
				\def\@p@sbbfile{\ps@founddir\ps@dir#1.bb}
			\fi
		\else
			\closein1
			\@decmprtrue
			\def\@p@sfile{#1}
			\def\@p@sbbfile{#1.bb}
		\fi
	\else
		\def\@p@sfile{\ps@founddir\ps@dir#1}
		\def\@p@sbbfile{\ps@founddir\ps@dir#1}
	\fi
   \else
	\closein1
	\def\@p@sfile{#1}
	\def\@p@sbbfile{#1}
   \fi
}
\def\@p@@sfile#1{\@p@@sfigure{#1}}
\def\@p@@sbbllx#1{
		\@bbllxtrue
		\dimen100=#1
		\edef\@p@sbbllx{\number\dimen100}
}
\def\@p@@sbblly#1{
		\@bbllytrue
		\dimen100=#1
		\edef\@p@sbblly{\number\dimen100}
}
\def\@p@@sbburx#1{
		\@bburxtrue
		\dimen100=#1
		\edef\@p@sbburx{\number\dimen100}
}
\def\@p@@sbbury#1{
		\@bburytrue
		\dimen100=#1
		\edef\@p@sbbury{\number\dimen100}
}
\def\@p@@sheight#1{
		\@heighttrue
		\dimen100=#1
   		\edef\@p@sheight{\number\dimen100}
}
\def\@p@@swidth#1{
		\@widthtrue
		\dimen100=#1
		\edef\@p@swidth{\number\dimen100}
}
\def\@p@@srheight#1{
		\@rheighttrue
		\dimen100=#1
		\edef\@p@srheight{\number\dimen100}
}
\def\@p@@srwidth#1{
		\@rwidthtrue
		\dimen100=#1
		\edef\@p@srwidth{\number\dimen100}
}
\def\@p@@sangle#1{
		\@angletrue
		\edef\@p@sangle{#1} 
}
\def\@p@@ssilent#1{ 
		\@verbosefalse
}
\def\@p@@sprolog#1{\@prologfiletrue\def\@prologfileval{#1}}
\def\@p@@spostlog#1{\@postlogfiletrue\def\@postlogfileval{#1}}
\def\@cs@name#1{\csname #1\endcsname}
\def\@setparms#1=#2,{\@cs@name{@p@@s#1}{#2}}
\def\ps@init@parms{
		\@bbllxfalse \@bbllyfalse
		\@bburxfalse \@bburyfalse
		\@heightfalse \@widthfalse
		\@rheightfalse \@rwidthfalse
		\def\@p@sbbllx{}\def\@p@sbblly{}
		\def\@p@sbburx{}\def\@p@sbbury{}
		\def\@p@sheight{}\def\@p@swidth{}
		\def\@p@srheight{}\def\@p@srwidth{}
		\def\@p@sangle{0}
		\def\@p@sfile{} \def\@p@sbbfile{}
		\def\@p@scost{10}
		\def\@sc{}
		\@prologfilefalse
		\@postlogfilefalse
		\@clipfalse
		\if@noisy
			\@verbosetrue
		\else
			\@verbosefalse
		\fi
}
\def\parse@ps@parms#1{
	 	\@psdo\@psfiga:=#1\do
		   {\expandafter\@setparms\@psfiga,}}
\newif\ifno@bb
\def\bb@missing{
	\if@verbose{
		\ps@typeout{psfig: searching \@p@sbbfile \space  for bounding box}
	}\fi
	\no@bbtrue
	\epsf@getbb{\@p@sbbfile}
        \ifno@bb \else \bb@cull\epsf@llx\epsf@lly\epsf@urx\epsf@ury\fi
}	
\def\bb@cull#1#2#3#4{
	\dimen100=#1 bp\edef\@p@sbbllx{\number\dimen100}
	\dimen100=#2 bp\edef\@p@sbblly{\number\dimen100}
	\dimen100=#3 bp\edef\@p@sbburx{\number\dimen100}
	\dimen100=#4 bp\edef\@p@sbbury{\number\dimen100}
	\no@bbfalse
}
\newdimen\p@intvaluex
\newdimen\p@intvaluey
\def\rotate@#1#2{{\dimen0=#1 sp\dimen1=#2 sp
		  \global\p@intvaluex=\cosine\dimen0
		  \dimen3=\sine\dimen1
		  \global\advance\p@intvaluex by -\dimen3
		  \global\p@intvaluey=\sine\dimen0
		  \dimen3=\cosine\dimen1
		  \global\advance\p@intvaluey by \dimen3
		  }}
\def\compute@bb{
		\no@bbfalse
		\if@bbllx \else \no@bbtrue \fi
		\if@bblly \else \no@bbtrue \fi
		\if@bburx \else \no@bbtrue \fi
		\if@bbury \else \no@bbtrue \fi
		\ifno@bb \bb@missing \fi
		\ifno@bb \ps@typeout{FATAL ERROR: no bb supplied or found}
			\no-bb-error
		\fi
		\count203=\@p@sbburx
		\count204=\@p@sbbury
		\advance\count203 by -\@p@sbbllx
		\advance\count204 by -\@p@sbblly
		\edef\ps@bbw{\number\count203}
		\edef\ps@bbh{\number\count204}
		\if@angle 
			\Sine{\@p@sangle}\Cosine{\@p@sangle}
	        	{\dimen100=\maxdimen\xdef\r@p@sbbllx{\number\dimen100}
					    \xdef\r@p@sbblly{\number\dimen100}
			                    \xdef\r@p@sbburx{-\number\dimen100}
					    \xdef\r@p@sbbury{-\number\dimen100}}
                        \def\minmaxtest{
			   \ifnum\number\p@intvaluex<\r@p@sbbllx
			      \xdef\r@p@sbbllx{\number\p@intvaluex}\fi
			   \ifnum\number\p@intvaluex>\r@p@sbburx
			      \xdef\r@p@sbburx{\number\p@intvaluex}\fi
			   \ifnum\number\p@intvaluey<\r@p@sbblly
			      \xdef\r@p@sbblly{\number\p@intvaluey}\fi
			   \ifnum\number\p@intvaluey>\r@p@sbbury
			      \xdef\r@p@sbbury{\number\p@intvaluey}\fi
			   }
			\rotate@{\@p@sbbllx}{\@p@sbblly}
			\minmaxtest
			\rotate@{\@p@sbbllx}{\@p@sbbury}
			\minmaxtest
			\rotate@{\@p@sbburx}{\@p@sbblly}
			\minmaxtest
			\rotate@{\@p@sbburx}{\@p@sbbury}
			\minmaxtest
			\edef\@p@sbbllx{\r@p@sbbllx}\edef\@p@sbblly{\r@p@sbblly}
			\edef\@p@sbburx{\r@p@sbburx}\edef\@p@sbbury{\r@p@sbbury}
		\fi
		\count203=\@p@sbburx
		\count204=\@p@sbbury
		\advance\count203 by -\@p@sbbllx
		\advance\count204 by -\@p@sbblly
		\edef\@bbw{\number\count203}
		\edef\@bbh{\number\count204}
}
\def\in@hundreds#1#2#3{\count240=#2 \count241=#3
		     \count100=\count240	
		     \divide\count100 by \count241
		     \count101=\count100
		     \multiply\count101 by \count241
		     \advance\count240 by -\count101
		     \multiply\count240 by 10
		     \count101=\count240	
		     \divide\count101 by \count241
		     \count102=\count101
		     \multiply\count102 by \count241
		     \advance\count240 by -\count102
		     \multiply\count240 by 10
		     \count102=\count240	
		     \divide\count102 by \count241
		     \count200=#1\count205=0
		     \count201=\count200
			\multiply\count201 by \count100
		 	\advance\count205 by \count201
		     \count201=\count200
			\divide\count201 by 10
			\multiply\count201 by \count101
			\advance\count205 by \count201
		     \count201=\count200
			\divide\count201 by 100
			\multiply\count201 by \count102
			\advance\count205 by \count201
		     \edef\@result{\number\count205}
}
\def\compute@wfromh{
		\in@hundreds{\@p@sheight}{\@bbw}{\@bbh}
		\edef\@p@swidth{\@result}
}
\def\compute@hfromw{
	        \in@hundreds{\@p@swidth}{\@bbh}{\@bbw}
		\edef\@p@sheight{\@result}
}
\def\compute@handw{
		\if@height 
			\if@width
			\else
				\compute@wfromh
			\fi
		\else 
			\if@width
				\compute@hfromw
			\else
				\edef\@p@sheight{\@bbh}
				\edef\@p@swidth{\@bbw}
			\fi
		\fi
}
\def\compute@resv{
		\if@rheight \else \edef\@p@srheight{\@p@sheight} \fi
		\if@rwidth \else \edef\@p@srwidth{\@p@swidth} \fi
}
\def\compute@sizes{
	\compute@bb
	\if@scalefirst\if@angle
	\if@width
	   \in@hundreds{\@p@swidth}{\@bbw}{\ps@bbw}
	   \edef\@p@swidth{\@result}
	\fi
	\if@height
	   \in@hundreds{\@p@sheight}{\@bbh}{\ps@bbh}
	   \edef\@p@sheight{\@result}
	\fi
	\fi\fi
	\compute@handw
	\compute@resv}
\def\OzTeXSpecials{
	\special{empty.ps /@isp {true} def}
	\special{empty.ps \@p@swidth \space \@p@sheight \space
			\@p@sbbllx \space \@p@sbblly \space
			\@p@sbburx \space \@p@sbbury \space
			startTexFig \space }
	\if@clip{
		\if@verbose{
			\ps@typeout{(clip)}
		}\fi
		\special{empty.ps doclip \space }
	}\fi
	\if@angle{
		\if@verbose{
			\ps@typeout{(rotate)}
		}\fi
		\special {empty.ps \@p@sangle \space rotate \space} 
	}\fi
	\if@prologfile
	    \special{\@prologfileval \space } \fi
	\if@decmpr{
		\if@verbose{
			\ps@typeout{psfig: Compression not available
			in OzTeX version \space }
		}\fi
	}\else{
		\if@verbose{
			\ps@typeout{psfig: including \@p@sfile \space }
		}\fi
		\special{epsf=\@p@sfile \space }
	}\fi
	\if@postlogfile
	    \special{\@postlogfileval \space } \fi
	\special{empty.ps /@isp {false} def}
}
\def\DvipsSpecials{
	\special{ps::[begin] 	\@p@swidth \space \@p@sheight \space
			\@p@sbbllx \space \@p@sbblly \space
			\@p@sbburx \space \@p@sbbury \space
			startTexFig \space }
	\if@clip{
		\if@verbose{
			\ps@typeout{(clip)}
		}\fi
		\special{ps:: doclip \space }
	}\fi
	\if@angle
		\if@verbose{
			\ps@typeout{(clip)}
		}\fi
		\special {ps:: \@p@sangle \space rotate \space} 
	\fi
	\if@prologfile
	    \special{ps: plotfile \@prologfileval \space } \fi
	\if@decmpr{
		\if@verbose{
			\ps@typeout{psfig: including \@p@sfile.Z \space }
		}\fi
		\special{ps: plotfile "`zcat \@p@sfile.Z" \space }
	}\else{
		\if@verbose{
			\ps@typeout{psfig: including \@p@sfile \space }
		}\fi
		\special{ps: plotfile \@p@sfile \space }
	}\fi
	\if@postlogfile
	    \special{ps: plotfile \@postlogfileval \space } \fi
	\special{ps::[end] endTexFig \space }
}
\def\psfig#1{\vbox {
	\ps@init@parms
	\parse@ps@parms{#1}
	\compute@sizes
	\ifnum\@p@scost<\@psdraft{
		\PsfigSpecials 
		\vbox to \@p@srheight sp{
			\hbox to \@p@srwidth sp{
				\hss
			}
		\vss
		}
	}\else{
		\if@draftbox{		
			\hbox{\fbox{\vbox to \@p@srheight sp{
			\vss
			\hbox to \@p@srwidth sp{ \hss 
			 \hss }
			\vss
			}}}
		}\else{
			\vbox to \@p@srheight sp{
			\vss
			\hbox to \@p@srwidth sp{\hss}
			\vss
			}
		}\fi

	}\fi
}}
\psfigRestoreAt
\setDriver
\let\@=\LaTeXAtSign

\begin{document}

\begin{center}

{\large \bf Detection of a Second Optical Component in the HI cloud
associated with the Young Dwarf Galaxy SBS~0335--052: New Data from
the 6--m Telescope.}

\vspace{0.5cm}

{\bf S.A.Pustilnik$^{1}$, V.A.Lipovetsky$^{1}$,Yu.I.Izotov$^{2}$,
E.Brinks$^{3}$, T.X.Thuan$^{4}$, A.Yu.Kniazev$^{1}$, S.I.Neizvestny$^{1}$,
A.V.Ugryumov$^{1}$}

\vspace{0.3cm}

$^{1}${\it Special Astrophysical Observatory,  Russian Academy of Sciences,
Nizhnij Arkhyz, Karachai--Circessia, 357147, Russia }

$^{2}${\it Main Astronomical Observatory, Goloseevo, Kiev, Ukraine}

$^{3}${\it NRAO, Socorro, NM, USA (currently on leave at the Departemento
de Astronom\' \i a, Universidad de Guanajuato, Guanajuato, M\'exico)}

$^{4}${\it Astronomy Department, University of Virginia, Charlottesville,
VA 22903, USA}

\vspace{0.7cm}

 Submitted to Astronomy Letters (Russia)

\end{center}

{\bf Abstract}

\vspace{0.5cm}

The Blue Compact Dwarf (BCD) galaxy SBS~0335--052 is one of the most
metal--deficient galaxies known and one of the best candidates for a
young dwarf galaxy in the process of formation.  A VLA HI map reveals
an unusual structure: the neutral gas is distributed in a very
extended disk, about 15 times larger than the Holmberg optical
diameter of the BCG.  There are two peaks of high density.  The
eastern peak is close to the position of \mbox{SBS~0335--052} whereas the
western HI peak is associated with a faint compact optical galaxy.  A
6--m telescope spectrum of this object shows H$\alpha$, H$\beta$, and
[OIII]$\lambda$5007 emission with a redshift close to that of
SBS~0335--052.  We suggest that it may be a young and chemically
unevolved dwarf galaxy.

\section {Introduction}

The blue compact galaxy SBS 0335--052 was discovered by Izotov et al.
(1990) as an extremely low metallicity galaxy, similar to I~Zw~18.
Its extremely low oxygen abundance [O/H]=7.26 has been confirmed by by
Terlevich et al. (1992) who used this galaxy for a new estimate of the
primordial helium abundance. An improved value of [O/H]=7.30 and an
NTT H$\alpha$ image was obtained by Melnick et al. (1992). They show
that the object has a double nuclear structure with a separation
between the two components of $\sim$1.0$''$.  The fainter component
has a higher abundance but it is still very metal poor, ([O/H]=7.64).
The H$\alpha$ image shows a complex structure with filaments and
wisps; the hydrogen lines extend to about 6$''$ on either side of the
nuclei.

Subsequent studies showed that SBS~0335--052 is an unusual HI gas--rich
dwarf galaxy (Thuan et al. 1996a), with no indications of an old
stellar population, either from high S/N MMT optical spectra (Izotov
et al. 1996), or from V and I imaging with HST (Thuan et al. 1995).
The images of SBS~0335--052 are generally reminiscent of I Zw 18
(Hunter and Thronson 1995). They show a large arc and several
star--forming regions in the central part surrounded by extended low
surface brightness emission of about 4 kpc in size, with systems of
loops typical for supernova related phenomena in star--forming
galaxies.  The age of these gas filaments is expected to be lower than
the age of the current star--formation burst (several million years).
Alternatively, the blue color of the extended halo could be due to
stellar light, but in this case, the stellar age also has to be less
than 100 million years.

The structure of the surrounding HI gas is especially interesting
since it can indicate how primeval galaxies collapse and form their
first stars.  We used the VLA in the C and D configurations to map and
study the velocity field of SBS~0335--052 in the 21--cm line.  The
observations were carried out in 1994--1995 for a total of 4.5 hours.
The VLA data will be discussed in detail in a forthcoming paper by
Thuan et al. (1996b).

\section {Optical identification}

When making the optical identifications we were guided by the
integrated HI flux density map.  The gas structure of this unique
young dwarf galaxy is quite unusual.  There are two prominent slightly
resolved HI peaks located at roughly equal distance from the center
of the HI cloud, separated by about 28~kpc. The location of the
eastern peak is close to the position of SBS~0335--052.

In Fig.1 we show the identification map extracted from the Digital Sky
Survey (DSS).  There is a faint, very compact, slightly elongated galaxy
at about 1.5$'$ West of SBS 0335--052. The observational parameters
for both eastern and western optical components are given in Table 1.
The position of the W component is very close to the coordinates of
the western HI peak and suggests that the two are physically
associated.  We estimate the integrated V luminosity of the W
component to be roughly 19$^m$. Its size as measured by its FWHM is
2.8$''\times$ 2.2$''$.

\section {Spectroscopy}

A spectrum of this western component (hereafter SBS~0335--052W) was
obtained with the 6--m telescope on December 17, 1996.  We used the
spectrograph SP--124 at the Nasmyth--1 focus in combination with a new
Tektronix 1024$\times$1024 CCD detector provided by the Astronomical
Institute at Potsdam (AIP) in the framework of a cooperative program
between SAO and the AIP.  Since the CCD detector had just been
installed for a test run, the instrument was not optimized and the
quality of the spectrum presented here is not representative of the
system at its best.  The weather conditions were not photometric with
fairly poor seeing.  A slit size of 35$''\times$2$''$ was used.  The
observations were carried out in a MIDAS environment, with the use of
the package NICE (Kniazev and Shergin 1995).

The spectrum covers the 4700$\div$7000 \AA \AA\ range with a resolution of
5.2 {\AA} pxl$^{-1}$.  Four 600 seconds exposures of variable quality were
averaged with appropriate weights to obtain the final spectrum.  The
spectrophotometric standards Hiltner~600 and Feige~34 were used to
correct the linearized spectrum for the spectral response of the total
system.  The reduction of the raw data was done in MIDAS with the
package LONG, adapted for our instrument by A.Ugryumov.

The final spectrum is shown in Fig.2.  Only three significant features
are seen in the spectrum, H$\alpha$, H$\beta$ and [OIII] 5007 \AA.\
Their equivalent widths are respectively 332, 93 and 98 \AA.  A very
faint feature is seen at the expected position of [OIII] 4959 \AA\
with an EW of about 25~\AA\ which is close to the expected value from the
theoretical ratio of the two [OIII] lines.  There is no indication of
either the [NII] or [SII] lines near H$\alpha$. For comparison we also
obtained the spectrum of SBS 0335--052 in the same night with a total
exposure time of 30 min.  The spectra were taken with a position angle
of --5$^\circ$ for the W component and of +10$^\circ$ for
SBS~0335--052.

It is interesting to estimate the extent of this W component. The
region which emits in H$\alpha$ can be traced on our long--slit spectra
over $\sim$13 pixels = 5.2$''$, the scale of the spectrum
perpendicular to the dispersion direction being 0.4$''$ pxl$^{-1}$.
This is about a factor of 3 less than the extent of H$\alpha$ in
SBS~0335--052 which covers about 40 pixels, as derived from the 6--m
telescope spectra obtained under the same conditions.

\section {Discussion}

The heliocentric velocities of 3990 km s$^{-1}$ for SBS~0335--052W,
and 4110 km s$^{-1}$ for SBS~0335--052 do not leave any doubt that
both objects belong to one system.  Both optical velocities coincide
within the errors with the velocities of the HI gas (see Table 1).
Thus in this unusual system we observe a pair of dwarf galaxies
embedded within a massive and huge HI cloud.  All the data indicate
that in the eastern HI density peak the formation of the galaxy
SBS~0335--052 started very recently. Its first massive stars were
formed during the last several million years.

\vspace{0.4cm}

 \begin{center}
Table 1. Observed and derived parameters for the SBS~0335--052 system

\vspace{0.4cm}

 \begin{tabular}{lcc} \hline \hline
 Parameter                      &      0335--052W    &     0335--052      \\
\hline

 $\alpha$(1950.0)$^a$ ...............................   &$03^h 35^m
09^s.57$&$03^h 35^m 15^s.15$    \\
 $\delta$(1950.0) .................................     &$-05^\circ12'24''.0$
&$-05^\circ12'25''.9$ \\
 V (mag) ..................................             &   19              &
 17                \\
 angular size ($''$)$^b$ ......................         &   5.2             &
 16                \\
 linear size (kpc)$^c$ .....................            &   1.3             &
 4.1               \\
 V$_{opt}$ (km s$^{-1}$) ..........................     & 3990$\pm$40       &
4110$\pm$20          \\
 V$_{HI}$  (km s$^{-1}$)$^d$ .........................  & 4006$\pm$\ 5      &
4068$\pm$\ 5          \\
 F(H$\alpha$)$\times$10$^{-16}$ (erg s$^{-1}$ cm$^{-2}$)...& ~~53$\pm$10 &
891$\pm$15            \\  \hline
 \end{tabular}
 \end{center}

{\small
\hspace{1.5cm}
$^a$ The coordinates were obtained from the DSS using MIDAS.

\hspace{1.5cm}
$^b$ The H$\alpha$ extension was used as the object's size.

\hspace{1.5cm}
$^c$ We adopted H$_\circ$=75 km s$^{-1}$ Mpc $^{-1}.$

\hspace{1.5cm}
$^d$ HI parameters (Thuan et al. 1996b).
}

\vspace{0.4cm}

The absolute magnitude of the western component is M$_{V}\sim$
-14.5$^m$ and its linear diameter is only about 1.3 kpc, taking the
H$\alpha$ extension as the size, which places it in the range of very
small blue compact dwarf galaxies. Its emission line spectrum
indicates that there is a current starburst in this galaxy as well.
No strong low--excitation line is seen in the red part of the spectrum
indicating, possibly, a high temperature and a low metal abundance.
The strong [OIII] and H$\beta$ lines seen do not contradict such a
possibility.  Our spectrum did not cover the region shortward of 4700
\AA, so we have no information on the range of possible temperatures
in the HII region. But if it turns out that its temperature is high
then the low intensity of [OIII] 5007 \AA\ relative to that of
H$\beta$ could be interpreted as an indication of a very low abundance
of oxygen. Clearly a high S/N spectrum of the western component is
necessary to elucidate its oxygen abundance and evolutionary status.

There are two possible models to explain this system.  Our preferred
hypothesis which takes into account all available data, is that the
eastern and western parts of SBS~0335--052 are two very young low
metallicity objects embedded in one huge HI cloud having approximately
the same young age.  A second scenario describing the nature of this
system is also possible, but seems to be not very probable. If the
metallicity of the western component turns out to indicate that it is
an older dwarf, we then witness an encounter of the dwarf with a
pristine huge HI--cloud. In that case the tidal disturbance of this
encounter has triggered the gas collapse in the eastern part of the HI
cloud and the subsequent dwarf galaxy formation (SBS~0335--052
itself).  Additional spectral observations and a detailed study of the HI
kinematics of this unique system will be necessary to make a choice between
these two models.

As for the possible tidal effect of another galaxy, indeed, such a
galaxy, the Scd spiral NGC~1376, is located at an angular distance of
about 9$'$ to the West of the SBS~0335--052 system. Its velocity,
determined from HI observations (Shostak 1975) is 4155$\pm$5 km
s$^{-1}$. The projected distance from the center of the SBS~0335--052
HI cloud is about 135 kpc.  However the role of this galaxy is far
from evident. If the radial velocities of SBS~0335--052 and NGC~1376
are due purely to Hubble flow, their relative distance is about 1500
kpc, and it would be hard to ascribe the very recent process of galaxy
formation in the SBS~0335--052 system to an influence exerted by
NGC~1376.

The only known BCD which is more metal--deficient than SBS~0335--052 is
I~Zw~18. It also has a double--component structure. But in I~Zw~18, the
two components embedded within an HI cloud are separated by only $\sim$
300 pc. This is very different from the situation in SBS~0335--052.
The newly discovered western component lies at a projected distance of
28 kpc and without the VLA map (Thuan et al. 1996b), we could not have
guessed that SBS~0335--052 and the its western component were parts of
one and the same physical system, spanning a total size 64 kpc.

Better knowledge of the kinematics of the neutral and ionized gas in
this key dwarf galaxy system in formation will allow us to shed some
light on the processes occurring at the epoch of galaxy formation, and
thus to predict better the properties of primeval galaxies.

\vspace{0.4cm}

The SAO authors are grateful to their German colleagues from the
Astrophysical Institute in Potsdam for the gift of a new CCD at the
6--m telescope which made possible these observations. Partial
financial support for this international collaboration was made
possible by NATO collaborative research grant 921285 and by INTAS
grant No 94--2285.

\vspace{0.5cm}

{\bf References}

\begin{enumerate}
\item Hunter D.A., Thronson H.A., 1995, Ap.J., {\bf 452}, 238
 \item Izotov Yu.I., Lipovetsky V.A., Guseva N.G., Kniazev A.Yu.,
Stepanian J.A., 1990, Nature, {\bf 343}, 238.
 \item Izotov Yu.I., Chaffee F., Foltz C., Guseva N.G., Lipovetsky V.A.,
   1996   (in preparation).
 \item Kniazev A.Yu., Shergin V.S., 1995, Special Astrophysical Obs.
   Technical Report No. {\bf 239}.
 \item Melnick J., Heydari--Malayeri M., Leisy P., 1992, A.Ap., {\bf 253}, 16.
 \item Shostak G.S., 1975, Ap.J., 189, L1.
 \item Terlevich E., Terlevich R., Skillman E., Stepanian J., Lipovetsky V.,
  1992, in Elements and Cosmos, eds.M.G.Edmunds, R.J.Terlevich, p.21,
Cambridge,
  University Press.
 \item Thuan T.X., Izotov Yu.I., Lipovetsky V.A., 1995, in Proc.of 11--th IAP
  Meeting, Paris.
 \item Thuan T.X., Lipovetsky V.A., Martin J.--M., Pustilnik S.A., 1996a, A.Ap.
  Suppl.Ser. (submitted).
 \item Thuan T.X., Brinks E., Pustilnik S.A., Lipovetsky V.A., Izotov Yu.I.,
  1996b (in preparation).

\end{enumerate}

\newpage
\begin{figure}
\centering
\vbox{

\psfig{figure=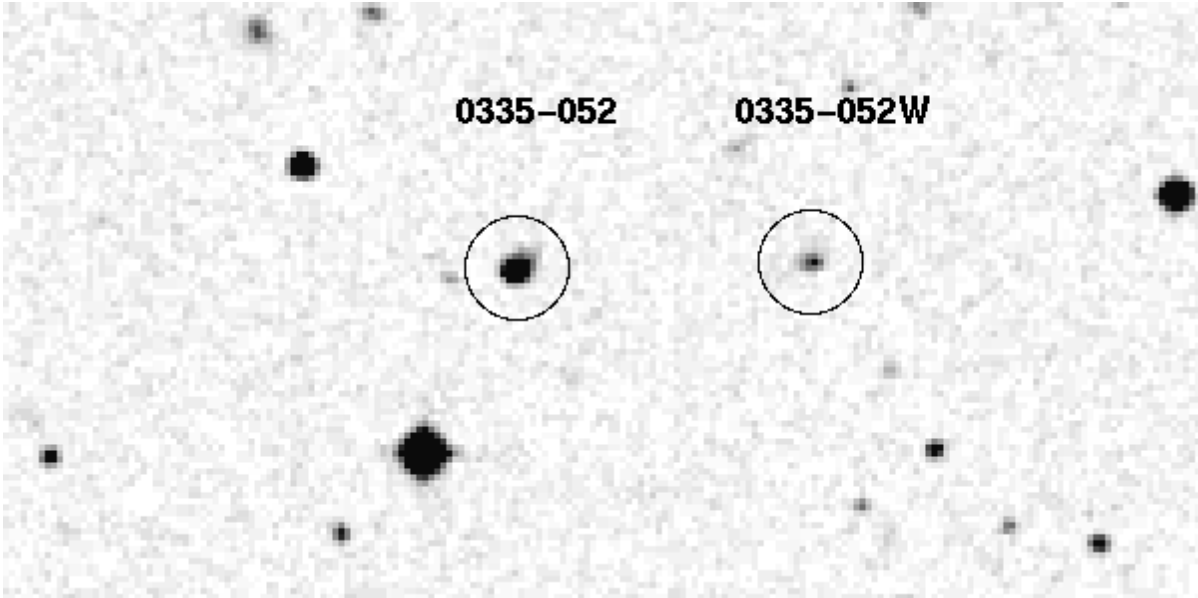,bbllx=70pt,bblly=500pt,bburx=522pt,bbury=755pt}\par}
\caption{
Identification chart from the DSS, 340$''\times$170$''$ in size.
SBS 0335--052 and the faint compact object 1.5$'$ to West are marked.}

\end{figure}

\begin{figure}
\centering
\vbox{

\psfig{figure=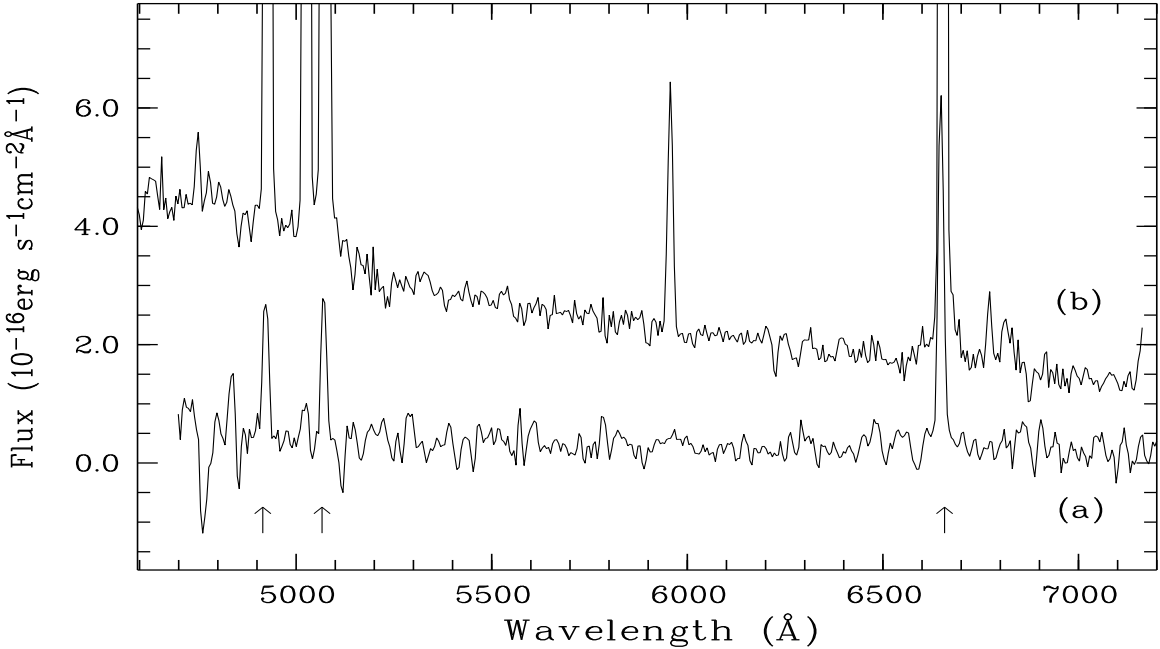,width=18cm,height=10cm,angle=270}\par}
\caption{
6--m telescope spectra of SBS~0335--052W (a) and SBS~0335--052 (b);
vertical arrows
show the position of emission lines of H$\beta$, [OIII] 5007 {\AA}, and
H$\alpha$.}

\end{figure}

\end{document}